\documentclass[prl,twocolumn,amsmath,amssymb,nofootinbib]{revtex4-1}
\usepackage{latexsym}
\usepackage{graphicx}
\usepackage{rotating}
\usepackage[normalem]{ulem}
\usepackage{hyperref}
\usepackage[version=4]{mhchem}
\usepackage{amsmath,amssymb,amsfonts}
\usepackage{bm}
\usepackage{color}
\usepackage{enumitem}
\usepackage{soul}

\usepackage{appendix}
\usepackage{graphicx}
\usepackage{epsfig}
\usepackage[justification=raggedright]{caption}
\usepackage{subfloat}
\usepackage[dvipsnames]{xcolor}
\usepackage[T1]{fontenc}
\usepackage{wrapfig}
\usepackage[english]{babel}
\usepackage{mathrsfs}
\usepackage{amsmath}
\usepackage{amsfonts}
\usepackage{sidecap}
\usepackage{bbm}
\usepackage[multidot]{grffile}
\usepackage{subfig}
\usepackage{multirow}
\usepackage{url}
\usepackage[english]{babel}
\usepackage{graphicx}
\usepackage{dcolumn}
\usepackage{bm}
\usepackage{times}
\usepackage{float}
\usepackage{physics}
\usepackage[small,compact]{titlesec}
\usepackage{hyperref}
\usepackage[utf8]{inputenc}
\usepackage[T1]{fontenc}
\usepackage{hyperref}
\usepackage{xr}
\newcommand{\be}{\begin{equation}}
\newcommand{\ee}{\end{equation}}

\makeatletter
\renewcommand*\env@matrix[1][*\c@MaxMatrixCols c]{%
  \hskip -\arraycolsep
  \let\@ifnextchar\new@ifnextchar
  \array{#1}}
\makeatother

\usepackage{times}
\begin{document}

\title{Electronic structure correspondence of singlet-triplet scale separation in strained \ce{Sr2RuO4}}
\author{Swagata Acharya}
\affiliation{ King's College London, Theory and Simulation of Condensed Matter,
              The Strand, WC2R 2LS London, UK}
\affiliation{Institute for Molecules and Materials, Radboud University, NL-6525 AJ Nijmegen, The Netherlands}
\author{Dimitar Pashov}
\affiliation{ King's College London, Theory and Simulation of Condensed Matter,
              The Strand, WC2R 2LS London, UK}
\author{Elena Chachkarova}
\affiliation{ King's College London, Theory and Simulation of Condensed Matter,
              The Strand, WC2R 2LS London, UK}
\author{Mark Van Schilfgaarde}
\affiliation{ King's College London, Theory and Simulation of Condensed Matter,
              The Strand, WC2R 2LS London, UK}
\affiliation{National Renewable Energy Laboratory, Golden, CO}
\author{C\'edric Weber}
\affiliation{ King's College London, Theory and Simulation of Condensed Matter,
              The Strand, WC2R 2LS London, UK}

\begin{abstract}

At a temperature of roughly 1\,K, \ce{Sr2RuO4} undergoes a transition from a normal Fermi liquid to a superconducting
phase.  Even while the former is relatively simple and well understood, the superconducting state is not even after 25
years of study.  More recently it has been found that critical temperatures can be enhanced by application of uniaxial
strain, up to a critical strain, after which it falls off.

In this work, we take an `instability' approach and seek for divergences in susceptibilities.  This provides an unbiased
way to distinguish tendencies to competing ground states.  We show that in the unstrained compound the singlet and
triplet instabilities of the normal Fermi liquid phase are closely spaced.  Under uniaxial strain electrons residing on
all orbitals contributing to the Fermiology become more coherent while the electrons of Ru-$d_{xy}$ character become
heavier and electrons of Ru-$d_{xz,yz}$ characters become lighter.  In the process, Im\,$\chi(\mathbf{q},\omega)$
increases rapidly around the incommensurate vector $\mathbf{q}{=}(0.3,0.3,0)2\pi/a$ while it gets suppressed at all
other commensurate vectors, in particular at $q{=}0$, which is essential for spin-triplet superconductivity.  Thus the
triplet superconducting instability remains the lagging instability of the system and the singlet instability enhances
under strain, leading to a large energy-scale separation between these competing instabilities.  At large strain an
instability to a spin density wave overtakes the superconducting one.

The analysis relies on a high-fidelity, \emph{ab initio} description of the one- particle properties and two-particle
susceptibilities, based on the Quasiparticle Self-Consistent \emph{GW} approximation augmented by Dynamical Mean Field
theory.  This approach is described and its high fidelity confirmed by comparing to observed one- and two-particle
properties. 

\end{abstract}

\maketitle


\section{Introduction}

The origin of superconducting pairing in \ce{Sr2RuO4} (SRO) has been one of the most debated topics in materials
research over last two decades~\cite{odder}.  Until recently the superconductivity was believed to be of spin-triplet
character.  A series of recent experimental findings, including strain dependent enhancement in the critical temperature
T$_{c}$~\cite{hicks,steppke} and the pronounced drop in O$^{17}$ NMR~\cite{pustogow} measurements, observation of
momentum-resolved superconducting energy gaps of \ce{Sr2RuO4} from quasiparticle interference imaging~\cite{arpes},
direct observation of Lifshitz transition~\cite{lifshitz}, jump in c$_{66}$ shear modulus~\cite{shearmodulus} and high
resolution $\mu$-SR studies~\cite{musr} have challenged the existing beliefs and demand a fresh look into the enigmatic
problem of superconductivity in SRO.

Strongly correlated electronic systems have a multiplicity of closely packed  phases, owing to
  the small energy scale of the different kinds of correlations.  Strain is an
effective tool to tune correlations in bulk crystalline systems, as it makes small but significant changes
  to the one-particle spectrum, which in turn modifies two-particle properties such as superconductivity.
It can lift degeneracies and separate out energy scales of competing phases, which sheds light
into the underlying mechanisms that lead to different orders. Sr$_{2}$RuO$_{4}$ is a particularly salient
  example: as noted a recent study showed that uniaxial strain induces a two-fold enhancement T$_{c}$ up to a critical
strain, after which it falls off rapidly~\cite{hicks,steppke}.  This study generated huge interest in the community and it was
followed by a series of careful experimental and theoretical works, including work by Steppke et
  al.~\cite{steppke} which attributed the increase to a van Hove singularity inducing a Lifshitz transition just around
  the critical point.  In the unstrained case, Sr$_{2}$RuO$_{4}$ has tetragonal symmetry, with three bands present at
  the Fermi level. These bands are composed predominantly of three Ru $d$ orbitals: the $d_{xy}$ and the
  symmetry-equivalent $d_{xz}$ and $d_{yz}$ pair.  Under strain the $d_{xz}$ and $d_{yz}$ equivalence is broken, and the
  Fermi surface undergoes a topological transition at a critical strain $\epsilon_x$ $\sim$ 0.6\%.

A series of theoretical studies~\cite{theory1,theory2,theory3,theory4} followed to explain the observations related to
enhancement and later suppression of T$_{c}$ under strain.  Some studies
rely on a starting electronic band structure from density functional theory (DFT); more often they are phenomenological
and based on low energy minimal model Hamiltonians.  The latter typically employ model parameters for the Hubbard
\emph{U} and \emph{J}, and often rely on DFT eigenvalues to parameterise the one-body part.  Such approaches
  are justified by the observation that superconductivity is a low energy phenomena, and should be well described if
  starting from a good underlying one-body part. Nevertheless, Kivelson et al., ~\cite{kivelson2020} recently argued
that while much is known about the normal phases of Sr$_{2}$RuO$_{4}$, understanding the nature of superconductivity in
\ce{Sr2RuO4} continues to be one of the most enigmatic problems in unconventional superconductivity even after 25
years~\cite{maeno94}. This is indeed a remarkable observation considering that the normal phase of \ce{Sr2RuO4} is a
relatively simple normal Fermi liquid, which is one of the better understood phases of correlated electronic materials.

In a recent work~\cite{swag19}, we performed a thorough analysis of \ce{Sr2RuO4} with and without uniaxial strain, using
a new high-fidelity \emph{ab-initio} approach~\cite{questaal_paper} to be described shortly.  It uses an instability
analysis: we monitor two-particle instabilities (points where a susceptibility diverges) in all particle-hole and
particle-particle channels, starting from high temperature and decreasing it.  This a significant departure from the
ground state low energy model Hamiltonian approach noted above, but we believe, it is key to addressing the right
questions for unconventional superconductivity, namely ``can we reliably compute all finite temperature instabilities in
the normal phase that on lowering of temperature would become unstable to a certain order?''  As Kivelson et al. noted,  
~\cite{kivelson2020} we believe, one key reason why superconductivity in \ce{Sr2RuO4} seems so difficult to explain stems from the
inability of theoretical schemes to calculate all possible two-particle instabilities in the normal phase.  This is
particularly difficult to accomplish in a parameter free fashion.  The instability analysis we use allows for
possible competing phases in an unbiased manner.  Further because the theory is both \emph{ab initio} and has very high
fidelity, it has unprecedented predictive power~\cite{nickel,swag19,Baldini,prl20,nemat}.  In this way we are able to
circumvent the difficulties Kivelson et al. noted.

Our \emph{ab initio} approach starts from a one-particle hamiltonian calculated from the quasiparticle self consistent
\emph{GW} (QS\emph{GW}) approximation~\cite{kotani}.  It plays the role of DFT as a bath for the many-body problem to
embedded in, but its fidelity is vastly superior.  The one-particle Green's function is generated from dynamical mean
field theory (DMFT)~\cite{georges1996}, using QS\emph{GW} as a bath.  This is accomplished with a Continuous Time
Quantum Monte Carlo (CTQMC) solver~\cite{hauleqmc,gull}.  This framework~\cite{prx,Baldini} is extended by computing the
local vertex from the two-particle Green's function by DMFT~\cite{hyowon_thesis,yin}, which is combined with nonlocal
bubble diagrams to construct a Bethe-Salpeter equation~\cite{swag19,prl20}.  The latter is solved to yield the essential
two-particle spin and charge susceptibilities $\chi^{d}$ and $\chi^{m}$ --- physical observables which provide an
important benchmark.  Moreover they supply ingredients needed for the Eliashberg equation, which yields eigenvalues and
eigenfunctions that describe instabilities to superconductivity in both singlet and triplet channels.  We will denote
QS\emph{GW}\textsuperscript{\footnotesize{++}} as a shorthand for the four-tier QS\emph{GW}+DMFT+BSE+Eliashberg theory.
The numerical implementation is discussed in Pashov et al.~\cite{questaal_paper} and codes are available on the open
source electron structure suite Questaal~\cite{questaal_web}.

QS\emph{GW}\textsuperscript{\footnotesize{++}} has high fidelity because QS\emph{GW} captures non-local dynamic
correlation particularly well in the charge channel~\cite{tomc, questaal_paper}, but it cannot adequately capture effects
of spin fluctuations.  DMFT does an excellent job at the latter, which are strong but mostly controlled by a local
effective interaction given by $U$ and $J$.  For \ce{Sr2RuO4} in particular the QS\emph{GW} Fermi surface is practically
indistinguishable from a recent high-resolution ARPES measurement~\cite{tamai2018}, and the spin susceptibility is in
excellent agreement with Inelastic Neutron Scattering (INS) measurements~\cite{swag19} (measured only for the unstrained
case when this work was published).

The present work reviews this prior study~\cite{swag19}, which was our first attempt to use instability analysis with
the full machinery of QS\emph{GW}\textsuperscript{\footnotesize{++}}.  It computed spin, charge and superconducting
susceptibilities resolved in both energy and momenta and both in the singlet and triplet channels.  We showed how the
singlet instability increases under strain, while the triplet one does not, and explained why T$_{c}$ increases.  Here
we extend that initial work to include wider excursion in strain to emphasize the trends, and provide a more detailed
description of the connection between the single-particle and two-particle properties.  In particular we establish the
following:

\begin{enumerate}[leftmargin=0cm]

\item show how strain modifies both one- and two-particle properties in a markedly orbital-dependent manner:
      strain enhances the role of the $d_{xy}$ orbital relative to the $d_{xz+yz}$ orbitals

\item show how the system becomes a better Fermi liquid with decreasing temperature.  At low temperature, \emph{J}
  becomes the dominant factor, and the increase in coherence is orbital specific, on account of the van
    Hove singularity

\item show how the system becomes a better Fermi liquid with increasing strain while at the same time $d_{xy}$ becomes
  heavier and $d_{xz}$ and $d_{yz}$ lighter.  Strain enhances the role of the $d_{xy}$ orbital relative to the
  $d_{xz+yz}$ orbitals, and enhances spin singlet superconductivity

\item Use instability analysis to clarify how the relative strength of competing phases evolve with strain, and compare
  against a spin density wave (SDW) (the latter eventually overtakes the instability towards superconductivity at a
  strain larger than the critical one)

\item Show how spin-orbit coupling affects superconductivity.

\end{enumerate}

In our original work we took \emph{U} and \emph{J} from constrained RPA~\cite{aryasetiawan} calculations
obtained from DFT~\cite{ujkotliar} which yielded \emph{U}=4.5\,eV and \emph{J}=1.0\,eV, and $J/U{=}0.22$. We have recently discovered from a recent implementation of C-RPA~\cite{questaal_web} that \emph{U} and \emph{J} computed from
DFT are too large to be used in a QS\emph{GW} framework: in the Hund's metals \emph{U} and $J$ decrease in proportion to
the bandwidth renormalisation, while $J/U$ remains fixed.  For \ce{Sr2RuO4} in particular, QS\emph{GW} renormalises the
DFT bandwidth by about 0.6.  Thus for the present study we use \emph{U}=3.0\,eV and \emph{J}=0.67\,eV; $J/U{=}0.22$.  This reduction
does not change anything qualitatively, but important details change, the most important being
that with the DFT-C-RPA estimates \emph{U} and \emph{J}, the leading triplet eigenvalue was found to be slightly larger than the singlet
eigenvalue in unstrained \ce{Sr2RuO4}, inconsistent with recent experimental
findings~\cite{arpes,musr,shearmodulus}.  In the present study we use the
newer parameterisation of \emph{U} and \emph{J}.

 Before turning to the results, we note that our original work emphasised the interplay between charge and
  spin susceptibility.  Those conclusions remain unchanged in the present work.  As we have nothing new to report on
  this aspect, we focus on analysis spin susceptibility, which we denote as $\chi(\mathbf{q},\omega)$, and label spin
  and charge susceptibilities as $\chi^m$ and $\chi^d$ only where a description of both is needed.  The superconducting
  instabilities we present here include both spin and charge susceptibilities.

\section{Results}


\noindent\emph{Single-particle properties near the Fermi surface:}

As noted earlier, the Fermi surface produced by QS\emph{GW} is essentially distinguishable from experiment (see SM,
Ref.~\cite{swag19}).  Augmentation with DMFT minimally affects the shape of the Fermi surface, but it does affect the
spin-orbit splitting.  QS\emph{GW}+DMFT yields 90-100\,meV, much larger than what had been widely thought, but in
excellent agreement with revised estimate of 100\,meV from a recent high-resolution laser ARPES
measurement~\cite{andrea1,tamai2018}. 

Further, the ability QS\emph{GW} or QS\emph{GW}+DMFT to yield a nearly perfect Fermi surface, to accurately predict the
spin-orbit splitting, and the critical $\epsilon_x$ $\sim$ 0.6\% where the Fermi surface undergoes a topological
transition, properties which DFT or DFT+DMFT do far less well, highlights the superior fidelity of QS\emph{GW}+DMFT.

\begin{figure*}
        \begin{center}
                \includegraphics[width=0.68\columnwidth]{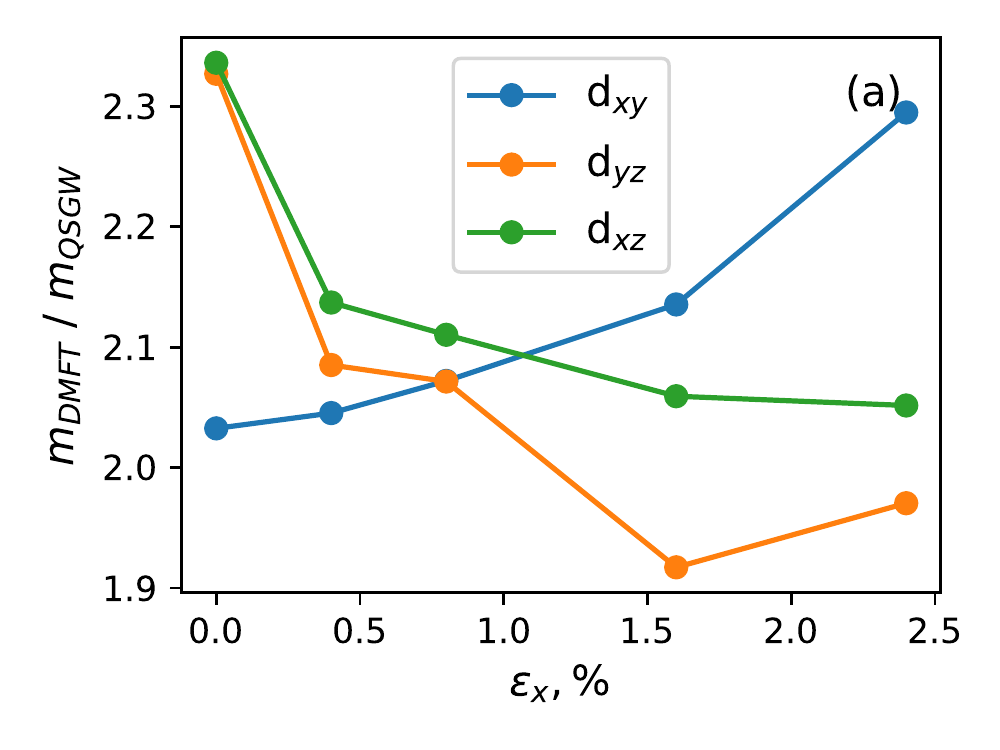}
                \includegraphics[width=0.68\columnwidth]{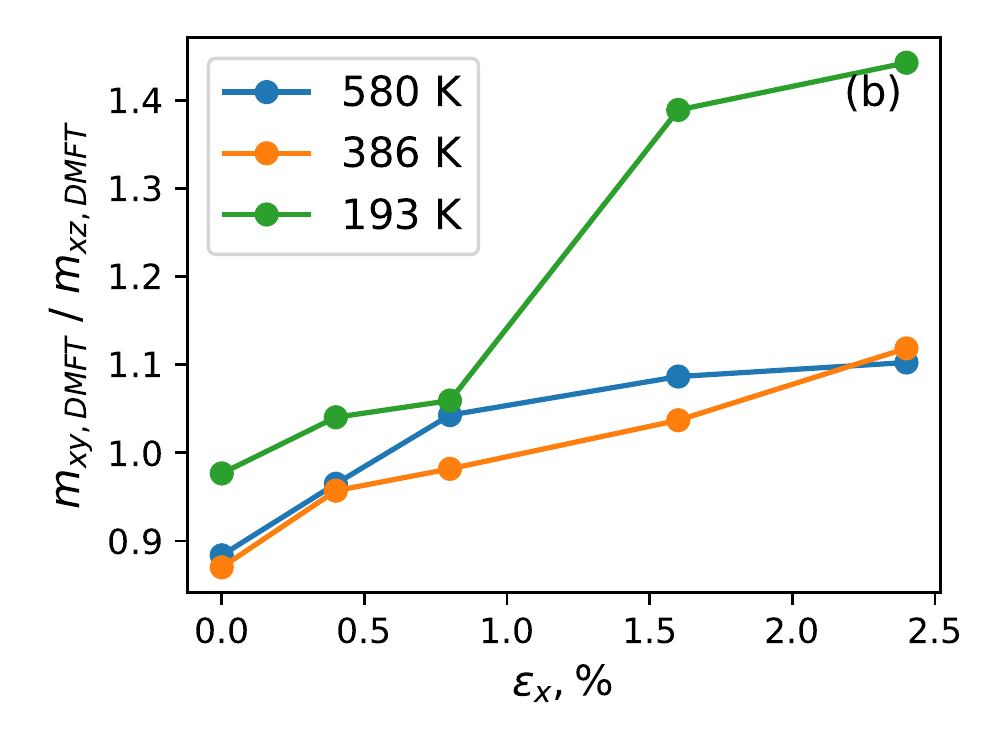}                
                 \includegraphics[width=0.68\columnwidth]{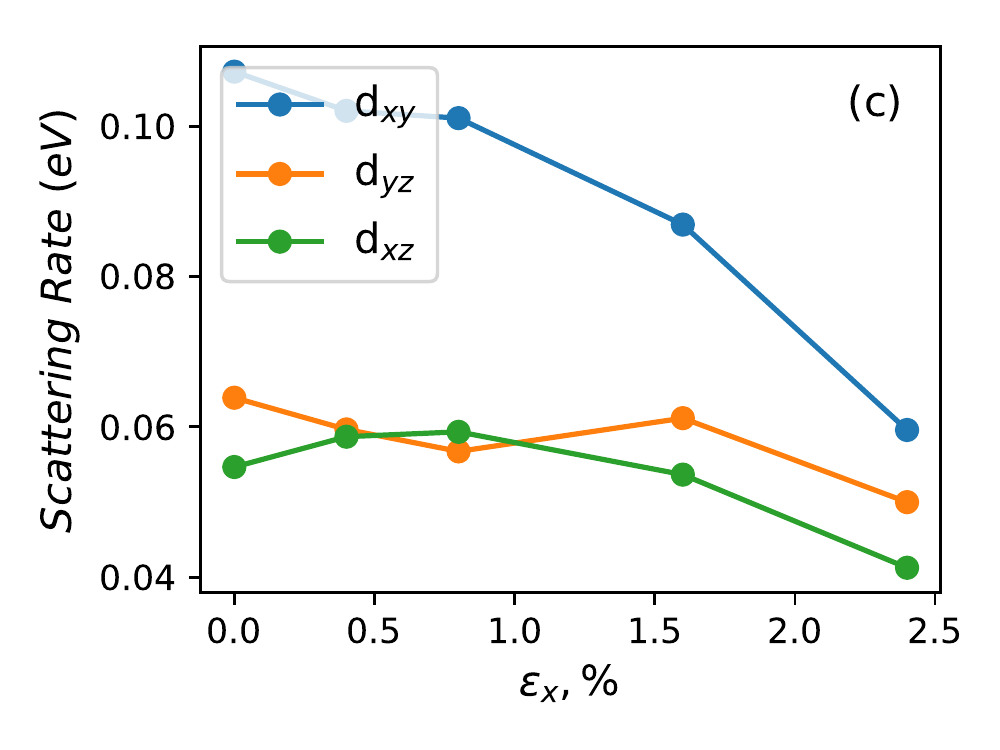}
                \caption{
                  {\bf Effective masses and scattering rates:} (a) The mass enhancement factors in DMFT (relative to the
                  QS\emph{GW} mass) are plotted in Ru-$d_{xy,yz,xz}$ channels. While the unstrained compound finds the
                  heaviest electron mass for the electrons in $d_{xz}$ orbital, under strain the $d_{xy}$ mass becomes
                  the heaviest. (b) We show the relative DMFT mass enhancement for d$_{xy}$ orbital in comparison to the d$_{xz}$ for all temperatures. (c) Scattering rates $\Gamma$ are orbitally
                  anisotropic, but under strain it decreases in all orbital channels. For very large strains the system
                  becomes a better Fermi liquid metal, nevertheless, the orbital anisotropy which is a typical signature
                  of Hund's metals survive for the entire range of strain.}
                \label{scatter}
        \end{center}
\end{figure*}

Fig.~\ref{scatter} (a) shows how the orbitally resolved electronic masses and single-particle scattering rates evolve with
strain $\epsilon_x$.  The single-site DMFT Im\,$\Sigma(i\omega)$ is fit to a fourth order polynomial in $i\omega$ for
low energies (first 6 Matsubara points at $\beta\,{=}\,40\,\text{eV}^{-1}{=}\,290$\,K)~\cite{nsrep}. The mass enhancement, related to
the coefficient $s_{1}$ of the linear term in the expansion
$m_{\text{DMFT}}/m_{\text{QS}GW}{=}1{+}|s_{1}|$~\cite{millis}, and the intercept $|s_{0}|$=$\Gamma
m_{\text{DMFT}}/m_{\text{QS}GW}$ with $m_{\text{DMFT}}$/$m_{\text{QS}GW}$ = $Z^{-1}$, is resolved in different
intra-orbital channels.  Both the masses and $\Gamma$ are orbital-dependent, and this differentiation is a signature of
a Hund's metal~\cite{prl20}.  Electrons in the $d_{xy}$ orbital become heavier while the $d_{xz,yz}$ electrons become
lighter with $\epsilon_x$. Beyond a critical $\epsilon_x$ $\sim$ 0.6\% 
$m_{xy,\text{DMFT}}$/$m_{xy,\text{QS}GW}$ becomes heavier than $m_{xz,\text{DMFT}}$/$m_{xz,\text{QS}GW}$ (see Fig.~\ref{scatter} (b)). The trend is similar at lower temperatures: the $d_{xy}$ mass increases
under strain while decreasing on $d_{xz}$ and $d_{yz}$. On the other
hand, all orbitals become more coherent under strain, as seen in the reduction of the scattering rate $\Gamma$ (see Fig.~\ref{scatter} (c)). 

\begin{figure*}
	\begin{center}
		\includegraphics[width=2.2\columnwidth]{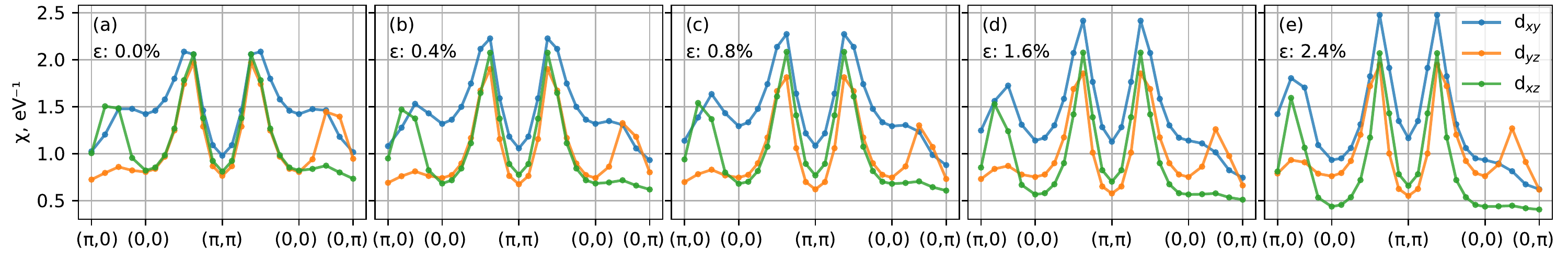}
			\caption{
			{\bf Orbital components of real part of static spin susceptibility Re$\chi(q,\omega=0)$:} We orbitally resolve the static spin susceptibility along some high-symmetry directions of the Brillouin Zone. The susceptibility at the ferromagnetic (FM) vector $\mathbf{q}^\mathrm{FM}{=}(0, 0, 0)$, $\chi$ is
			dominated by the intra-orbital fluctuations in the $d_{xy}$ channel, while at the incommensurate (IC) vector
			$\mathbf{q}^\mathrm{IC}{=}(0.3, 0.3, 0)$ (we use units $2\pi/a$ throughout) the three orbitals contribute almost
			equally. The antiferromagnetic (AFM) vector $\mathbf{q}^\mathrm{AFM}{=}(0.5, 0.5, 0)$ is fully gapped. 
			Under strain the IC peak rapidly increases, and $d_{xy}$ emerges as the leading component of total spin susceptibilities along all high symmetry directions.
			}
		\label{rechi}
	\end{center}
\end{figure*}

\noindent\emph{Spin fluctuations: incommensurability and coherence:}

$\chi(\mathbf{q},\omega)$ is computed from the momentum dependent Bethe-Salpeter equations~\ref{eq:BSE_imag} in the magnetic channel.
\begin{equation}
\chi_{{\alpha_{1},\alpha_{2}\atop \alpha_{3},\alpha_{4}}}^{m}(i\nu,i\nu^{\prime})_{\mathbf{q},i\omega}=
[(\chi^{0})_{\mathbf{q},i\omega}^{-1}-\Gamma_{loc}^{irr,m}]_{{\alpha_{1},\alpha_{2}\atop \alpha_{3},\alpha_{4}}}^{-1}(i\nu,i\nu^{\prime})_{\mathbf{q},i\omega}.
\label{eq:BSE_imag}
\end{equation}

\begin{figure*}
\includegraphics[width=2.0\columnwidth]{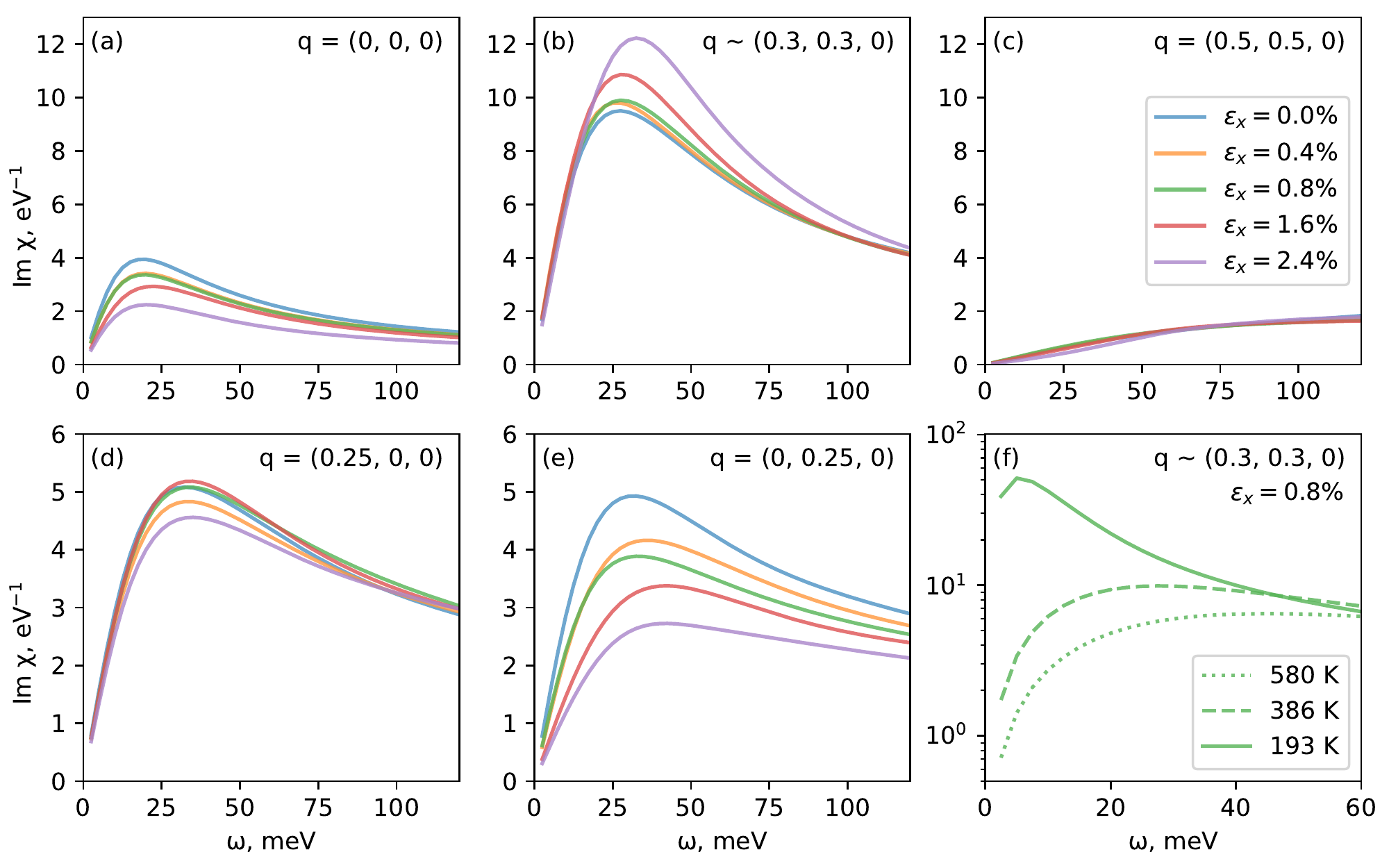}
 \caption{{\bf Strain and temperature dependence of susceptibilities:} (a)-(e) Imaginary part of the dynamic spin susceptibility
   $\chi(\mathbf{q},\omega)$ at some high symmetry points in the Brillouin zone for different strains $\epsilon_x$.
   The unstrained compound shows a spin fluctuation spectrum strongly peaked at $(0.3,0.3,0)$. With
   increasing strain fluctuations become more strongly peaked at $(0.3,0.3,0)$ while it gets suppressed at the
   ferromagnetic vector and remains fully gapped at the anti-ferromagnetic vector. (f) With lowering temperature the IC peak at q=$(0.3,0.3,0)$ start to diverge for strains  $\epsilon_x >$0.6\%, signaling an instability towards a a spin density wave order. }
\label{imchi}
\end{figure*}

$\chi^{0}$ is the non-local ($k$-dependent) polarisation bubble computed from single-particle QS\emph{GW} Green's
functions dressed by the local DMFT self-energy, and $\Gamma$ is the local irreducible two-particle vertex function
computed in the magnetic channel.  $\Gamma$ is a function of two fermionic frequencies $\nu$ and $\nu'$ and the bosonic
frequency $\omega$.  $\chi(\mathbf{q},i\omega)$ is computed by closing
$\chi_{{\alpha_{1},\alpha_{2}\atop \alpha_{1},\alpha_{2}}}^{m(d)}(i\nu,i\nu^{\prime})_{\mathbf{q},i\omega}$ with spin
bare vertex $\gamma$ and summing over frequencies ($i\nu$,$i\nu^{\prime}$) and orbitals ($\alpha_{1,2}$).

We compute the real part of the static susceptibility $\chi(\mathbf{q},i\omega{=}0)$ and resolve it in different inter-
and intra-orbital channels to develop a systematic understanding of which orbitals dominate the spin susceptibilities at
different q-vectors.  In the vicinity of the ferromagnetic (FM) vector  $\mathbf{q}^\mathrm{FM}{=}(0, 0, 0)$, $\chi$ is
dominated by the intra-orbital fluctuations in the $d_{xy}$ channel (Fig.~\ref{rechi} (a)), while at the incommensurate (IC) vector
$\mathbf{q}^\mathrm{IC}{=}(0.3, 0.3, 0)$ (we use units $2\pi/a$ throughout) the three orbitals contribute almost
equally. The antiferromagnetic (AFM) vector $\mathbf{q}^\mathrm{AFM}{=}(0.5, 0.5, 0)$ is fully gapped. 

When strain is applied we find that the IC peak rapidly increases, and $d_{xy}$ emerges as the leading component of
total spin susceptibilities along all high symmetry directions (see Fig.~\ref{rechi} (b-e)). This is consistent with the fact that under strain,
$d_{xy}$ becomes the most strongly correlated orbital. Nevertheless, the AFM vector remains fully gapped for strains up
to $\epsilon_x$=2.4\%.  We compute both real and imaginary parts of spin and charge susceptibilities by solving the BSE
in respective channels. These equations are solved in the Matsubara representation with local dynamic vertex functions
(which are functions of three Matsubara frequencies) and the non-local polarisation bubble which also has the Matsubara
frequencies.  After summing over all internal Fermionic Matsubara frequencies and orbital indices, we are left with
$\chi(\mathbf{q},i\omega)$. Further, it needs to be analytically continued to real bosonic frequencies.  One way is to
analytically continue $\chi(i\omega)$ at each momentum, which is tremendously expensive.  To understand the precise
nature of the spin fluctuations at finite energies, it was imperative in this work that we extract
Im\,$\chi(\mathbf{q},\omega)$ for finite $\omega$. For low energies, which is the focus here, the vertex
$\Gamma^{irr}_{loc}$ is analytically continued by a quasiparticle-like approximation. We replace the frequency-dependent
vertex with a constant, i.e., $\Gamma^{irr}_{loc}(i\nu,i\nu',i\omega)_{{\alpha_{2} \sigma_{2},\alpha_{4} \sigma_{4}
    \atop \alpha_{1} \sigma_{1},\alpha_{3} \sigma_{3}}} \sim U^\text{eff}_{{\alpha_{2} \sigma_{2},\alpha_{4} \sigma_{4}
    \atop \alpha_{1} \sigma_{1},\alpha_{3} \sigma_{3}}}$ which satisfies the constraint that
$\chi(\mathbf{q},i\omega{=}0)=\chi(\mathbf{q},\omega{=}0)$. 

This ``quasiparticlized'' vertex $U^\text{eff}$ contains all the important spin, orbital dependence. This approximation for
analytic continuation works remarkably well for spin susceptibilities at low energy as shown in previous
works~\cite{yin,hyowon_thesis,prl20,swag19}.  We compute the dynamic susceptibility Im\,$\chi(\mathbf{q},\omega)$ and observe
that the intensity drops at $\mathbf{q}^\mathrm{FM}{=}(0, 0, 0)$ under strain (see Fig.~\ref{imchi} (a)). The energy dispersion of Im\,$\chi(\mathbf{q},\omega)$ at
$\mathbf{q}^\mathrm{FM}{=}(0, 0, 0)$ remains almost invariant up to $\epsilon_x$=1.6\%, but for much larger strains the branch
loses both intensity and dispersion simultaneously. The reverse happens at $\mathbf{q}^\mathrm{IC}{=}(0.3, 0.3, 0)$ where under
strain both the intensity and dispersion of the branch increase (see Fig.~\ref{imchi} (b)). For all strains, the $\mathbf{q}^\mathrm{AFM}{=}(0.5, 0.5, 0)$
remains fully gapped (see Fig.~\ref{imchi} (c)), while the peaks at $\mathbf{q}{=}(0.25, 0, 0)$ and $\mathbf{q}{=}(0, 0.25, 0)$ lose intensity, but in a very
anisotropic manner (see Fig.~\ref{imchi} (d-e)). We also show in  Fig.~\ref{imchi} (f) how the IC peak starts diverging with lowering temperatures at $\epsilon_x$=0.8\%, signaling an instability towards an SDW order. However, whether the Fermi liquid phase will become unstable to an SDW phase or a superconducting phase, can only be confirmed from further investigation of superconducting pairing instabilities. 

\begin{figure*}

                \includegraphics[width=1.0\columnwidth]{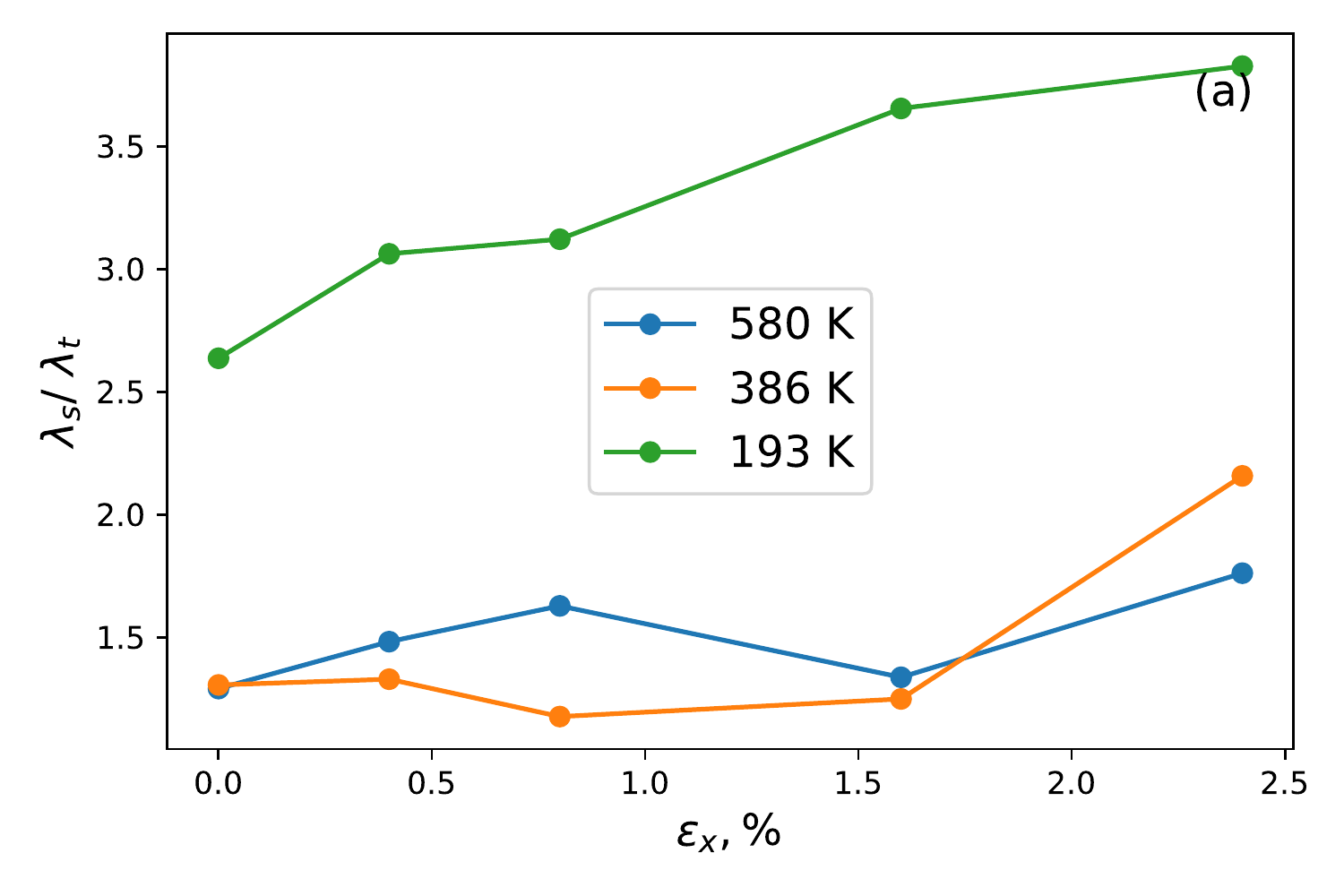}
                \includegraphics[width=1.0\columnwidth]{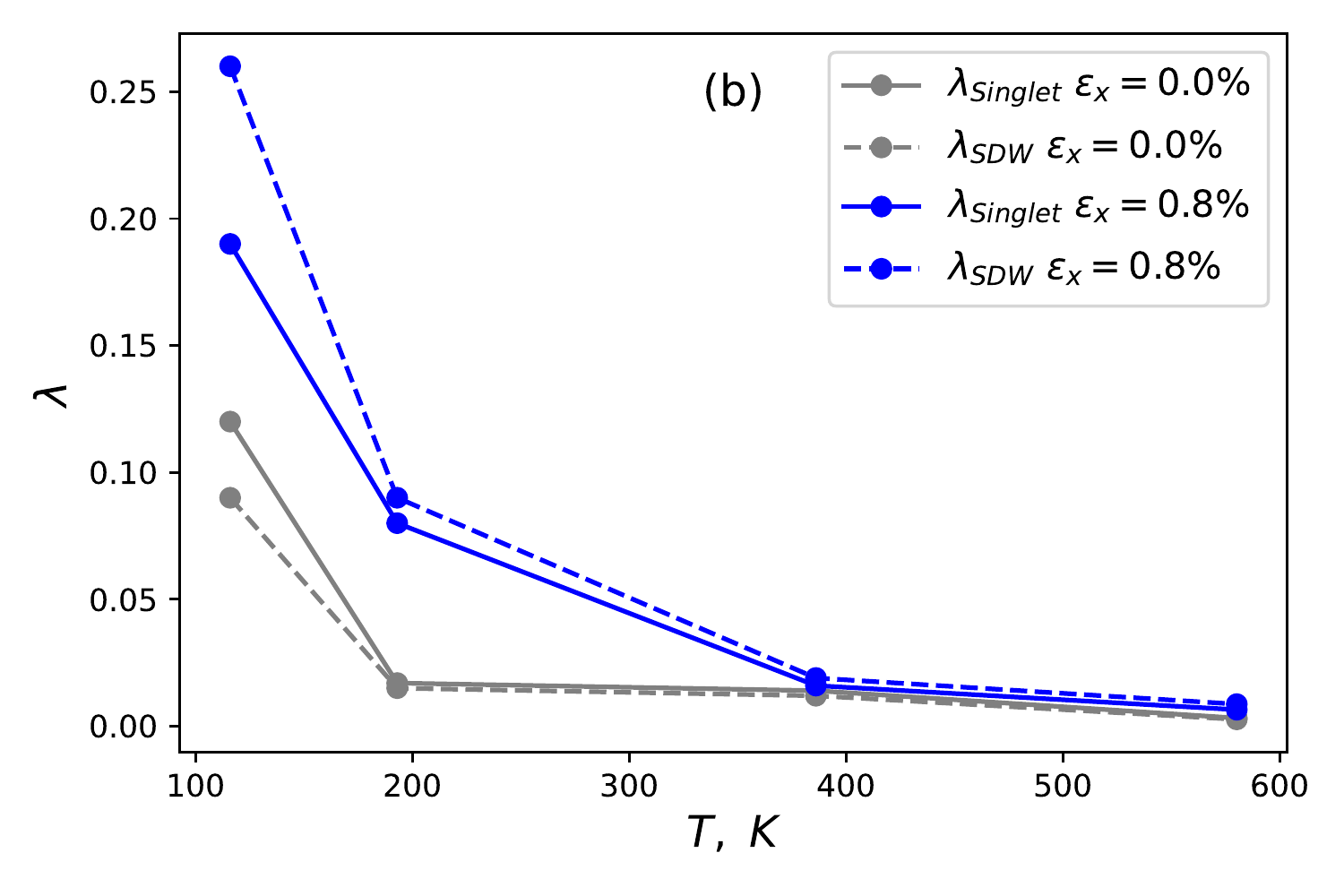}
                \caption{{\bf Superconducting eigenvalues: singlet-triplet scale separation and SDW}: (a) We plot the relative
                  strength of the leading singlet eigenvalue in comparison the triplet eigenvalue
                  ($\lambda_{s}/\lambda_{t}$) extracted by solving the multi-orbital Bethe Salpeter equation in the
                  superconducting channels (both singlet and triplet) as functions of temperature. With strain
                  $\lambda_{s}/\lambda_{t}$ increases and the trend becomes more prominent with decreasing temperature. (b) In the unstrained compound the leading eigenvalue ($\lambda_{s}$) singlet superconducting instability has slightly steeper temperature dependence in comparison to the eigenvalue ($\lambda_{SDW}$) for SDW instability. However, under strain, beyond $\epsilon_x$=0.6\%, the SDW instability becomes the leading instability of the system on lowering temperatures.}
                \label{sc}

\end{figure*}



\noindent\emph{Superconducting pairing: nodal character and dimensionality:}

The superconducting pairing susceptibility $\chi^{p-p}$ is computed by dressing the non-local pairing polarisation
bubble $\chi^{0,p-p}(k,i\nu)$ with the pairing vertex $\Gamma^{irr,p-p}$ using the Bethe-Salpeter equation in
the particle-particle channel (see Supplementary Figure 5 for Feynman diagram representation).
\begin{eqnarray}
\chi^{p-p} = \chi^{0,p-p}\cdot[\mathbf{1}+\Gamma^{irr,p-p}\cdot\chi^{0,p-p}]^{-1}
\end{eqnarray}
$\Gamma^{irr,p-p}$ in the singlet (s) and triplet (t) channels are obtained from the magnetic (spin) and density (charge) particle-hole reducible vertices by
\begin{eqnarray} 
&\Gamma_{{\alpha_{2},\alpha_{4}\atop \alpha_{1},\alpha_{3}}}^{irr,p-p,s}(\textbf{k},i\nu,\textbf{k}',i\nu') = \Gamma_{{\alpha_{2},\alpha_{4}\atop \alpha_{1},\alpha_{3}}}^{f-irr}(i\nu,i\nu')\nonumber\\
&+\frac{1}{2}[\frac{3}{2}\widetilde{\Gamma}^{p-h,(m)}\nonumber\\
&-\frac{1}{2}\widetilde{\Gamma}^{p-h,(d)}]_{{\alpha_{2},\alpha_{3}\atop \alpha_{1},\alpha_{4}}}(i\nu,-i\nu')_{\textbf{k}'-\textbf{k},i\nu'-i\nu}\nonumber\\
&+\frac{1}{2}[\frac{3}{2}\widetilde{\Gamma}^{p-h,(m)}\nonumber\\
&-\frac{1}{2}\widetilde{\Gamma}^{p-h,(d)}]_{{\alpha_{4},\alpha_{3}\atop \alpha_{1},\alpha_{2}}}(i\nu,i\nu')_{-\textbf{k}'-\textbf{k},-i\nu'-i\nu}
\label{eq:Gamma_pp_nonloc1}
\end{eqnarray}

\begin{eqnarray}
&\Gamma_{{\alpha_{2},\alpha_{4}\atop \alpha_{1},\alpha_{3}}}^{irr,p-p,t}(\textbf{k},i\nu,\textbf{k}',i\nu') = \Gamma_{{\alpha_{2},\alpha_{4}\atop \alpha_{1},\alpha_{3}}}^{f-irr}(i\nu,i\nu')\nonumber\\
&-\frac{1}{2}[\frac{1}{2}\widetilde{\Gamma}^{p-h,(m)}\nonumber\\
&+\frac{1}{2}\widetilde{\Gamma}^{p-h,(d)}]_{{\alpha_{2},\alpha_{3}\atop \alpha_{1},\alpha_{4}}}(i\nu,-i\nu')_{\textbf{k}'-\textbf{k},i\nu'-i\nu}\nonumber\\
&+\frac{1}{2}[\frac{1}{2}\widetilde{\Gamma}^{p-h,(m)}\nonumber\\
&+\frac{1}{2}\widetilde{\Gamma}^{p-h,(d)}]_{{\alpha_{4},\alpha_{3}\atop\alpha_{1},\alpha_{2}}}(i\nu,i\nu')_{-\textbf{k}'-\textbf{k},-i\nu'-i\nu}
\label{eq:Gamma_pp_nonloc2}
\end{eqnarray}
Finally, $\chi^{p-p}$ can be represented in terms of eigenvalues $\lambda$ and eigenfunctions $\phi^{\lambda}$
of the Hermitian particle-particle pairing matrix (see the SM for detailed derivation).
\begin{eqnarray}
\chi^{p-p}(k,k') & = & \sum_{\lambda}\frac{1}{1-\lambda}\cdot(\sqrt{\chi^{0,p-p}(k)}\cdot\phi^{\lambda}(k))\nonumber\\
&\cdot(&\sqrt{\chi^{0,p-p}(k')}\cdot\phi^{\lambda}(k'))
\end{eqnarray}
The pairing susceptibility diverges when the leading eigenvalue $\lambda$ becomes unity. The corresponding eigenfunction
represents the momentum structure of $\chi^{p-p}$.  Unconventional superconductivity in SRO is multi-orbital in nature
with multiple competing instabilities. In our previous work~\cite{swag19}, we performed a thorough analysis of all
possible singlet and triplet instabilities in SRO and associated with a particular symmetry group.  We showed that the
leading eigenvalue in the singlet channel had a $d_{x^2-y^2}$ instability (B$_{1g}$ symmetry) while the leading
eigenvalue in the triplet channel was of an extended nodeless \emph{s}-wave 2$\delta_{0}+\cos k_{x}+\cos k_{y}$ gap
structure with A$_{1g}$ irreducible representation in the $d_{xz,yz}$ basis.

A subsequent Bogoliubov quasiparticle scattering interference visualization of the gap structure at milli-Kelvin
temperatures was measured to be of B$_{1g}$-$d_{x^2-y^2}$ nature.~\cite{arpes} We observe that for all strains (and
without strain) the eigenvalue corresponding to the singlet instability remains the leading one and the relative
strength of the singlet to triplet eigenvalues ($\lambda_{s}/\lambda_{t}$) keep increasing under strain.  The
enhancement in $\lambda_{s}/\lambda_{t}$ under strain, becomes more apparent at lower temperatures (see Fig.~\ref{sc} (a))).  This is concomitant
with the mass becoming heavier in the $d_{xy}$ channel while the masses relax on other orbitals.  Further, this is a
direct consequence of the spin fluctuations getting suppressed at $\mathbf{q}^\mathrm{FM}{=}(0, 0, 0)$ and rising steeply at
$\mathbf{q}^\mathrm{IC}{=}(0.3, 0.3, 0)$.  It is understandable that the system can undergo a spin density wave order mediated
primarily via the fluctuations at and around $\mathbf{q}^\mathrm{IC}{=}(0.3, 0.3, 0)$.  Once the spin susceptibility diverges, at
lower temperatures, under large strains, the system will encounter the density wave phase and the superconducting
channel will be suppressed.  To check that we extract the leading eigenvalue ($\lambda_{SDW}$) in the density wave
channel, by diagonalising the susceptibility matrix.  We observe that while for $\epsilon_x$=0.0, the $\lambda_{SDW}$
and $\lambda_{s}$ show a very similar temperature dependence ($\lambda_{s}$ is slightly more steeper than $\lambda_{SDW}$), for finite and large strains
($\epsilon_x{>}{0.6\%}$) $\lambda_{SDW}$ acquires a steeper temperature dependence than $\lambda_{s}$ (see Fig.~\ref{sc} (b)). This
suggests that although the $\lambda_{s}/\lambda_{t}$ continues to enhance under large strains, the superconducting phase
will be suppressed by a SDW phase : the normal Fermi liquid phase will make a transition to the SDW phase before it
becomes superconducting~\cite{sdw,steppke}. 

We observe that all our essential conclusions for both spin and superconducting instabilities remain qualitatively invariant once the spin-orbit coupling (SOC) is included in the calculations. We observe that under strain, with SOC the singlet and triplet eigenvalues get further removed from each other, making the scale separation clearer for all strains. The FM spin fluctuations go down under strain, in presence of SOC and the IC becomes steeper, making a SDW instability likely for larger strains.

\noindent\emph{Summary}

We have performed a detailed analysis of the single-particle and two-particle response of \ce{Sr2RuO4} under large
strains.  The instability approach allows us to compare different kinds of instabilities of the normal phase.  By
performing excursions in temperature or external parameters such as strain we can identify which ground states are
preferred instabilities of the normal phase, distinguishing among multiple closely spaced many-body ordered phases.  Key
to the success of this approach is the \emph{ab initio} QS\emph{GW}\textsuperscript{\footnotesize{++}} machinery, whose
high fidelity (which is essential) is confirmed by the excellent agreement with observed one- and two-particle
properties, as we have shown.

We find that while the singlet and triplet instabilities are similar in the unstrained \ce{Sr2RuO4}, the ratio of
eigenvalues $\lambda_{s}/\lambda_{t}$ under uniaxial strain $\epsilon_x$ keeps increasing at all temperatures, leading
to a clear separation between the singlet and the triplet superconducting pairing instabilities. Its emergence can be
traced to the orbital-selective evolution in single-particle properties under strain: particularly $d_{xy}$ acquires a
heavy mass while $d_{xz}$ and $d_{yz}$ become lighter.  This directly modifies the two-particle susceptibilities; the
spin susceptibility at $\mathbf{q}^\mathrm{FM}$ is suppressed under strain and at $\mathbf{q}^\mathrm{IC}$ it diverges,
leading to the relative suppression of the triplet instability.  Finally, the rapid divergence of $\chi$ temperature at
$\mathbf{q}^\mathrm{IC}$ leads to enhancement in both $\lambda_{s}$ and $\lambda_{SDW}$.  The latter has a steeper
temperature dependence, and thereby, for large strains the superconducting phase is suppressed by an emergent SDW phase.

\section*{Methods}

We use a recently developed quasi-particle self consistent \emph{GW} + dynamical mean field theory
(QS\emph{GW}+DMFT)~\cite{prx,nickel,questaal_paper}, as implemented in the all-electron Questaal package \cite{questaal_web}. Paramagnetic DMFT is
combined with nonmagnetic QS\emph{GW} via local projectors of the Ru 4$d$ states on the Ru augmentation spheres to form
the correlated subspace.  We carried out the QS\emph{GW} calculations in the tetragonal and strained phases of
\ce{Sr2RuO4} with space group 139/I4mmm.  DMFT provides a non-perturbative treatment of the local spin and charge
fluctuations. We use an exact hybridisation expansion solver, namely the continuous time Monte Carlo (CTQMC)
\cite{Haule_long_paper_CTQMC}, to solve the Anderson impurity problem.

The one-body part of QS\emph{GW} is performed on a $16 \times 16 \times 16$ k-mesh and charge has
been converged up to $10^{-6}$ accuracy, while the (relatively smooth) many-body static self-energy $\Sigma^0(\mathbf{k})$
is constructed on a $8 \times 8 \times 8$ k-mesh from the dynamical $GW$ $\Sigma(\mathbf{k},\omega)$.
$\Sigma^0(\mathbf{k})$ is iterated until convergence (RMS change in $\Sigma^0{<}10^{-5}$\,Ry).
$U$=3.0\,eV and $J$=0.67\,eV  were used as correlation parameters for DMFT. The DMFT for the dynamical self energy is iterated, and converges in $\approx 20$ iterations.
Calculations for the single particle response functions are performed with $10^9$ QMC steps per core and the statistics
is averaged over 64 cores. The two particle Green's functions are sampled over a larger number of
cores (10000-20000) to improve the statistical error bars. We sample the local two-particle Green's functions
with CTQMC for all the correlated orbitals and compute the local polarisation
bubble to solve the inverse Bethe-Salpeter equation (BSE) for the local
irreducible vertex.
Finally, we compute the non-local polarisation bubble $G(\mathbf{k},\omega)G(\mathbf{k}{-}\mathbf{q},\omega{-}\Omega)$
and combined with the local irreducible vertex \cite{hywon_vertex} we obtain the full non-local spin and charge
susceptibilities $\chi^{m,d}(\mathbf{q},\omega)$. The susceptibilities are computed on a $16 \times 16 \times 16$
$\mathbf{Q}$-mesh. BSE equations in the particle-particle pairing channels are solved~\cite{swag19,prl20} on the same k-mesh to extract the
susceptibilities and the Eliashberg eigenvalue equations are solved to extract the eigenvalue spectrum and corresponding
pairing symmetries.

\section*{Acknowledgments}
SA acknowledges discussions with Stephen Hayden, James Annett, Seamus Davis and Astrid Romer. This work was supported by the Simons Many-Electron Collaboration.
For computational resources, MvS, SA and DP acknowledge PRACE for awarding us access to SuperMUC at GCS@LRZ, Germany and Irene-Rome hosted by TGCC, France.



%

\end{document}